\documentclass[aps,prb,twocolumn,showpacs,floats]{revtex4}
\usepackage{graphicx}%
\usepackage{epsfig}
\usepackage{amsmath}%
\usepackage{amssymb}%

\def\beq{\begin{equation}}

\def\eeq{\end{equation}}
\def\beqa{\begin{eqnarray}}
\def\eeqa{\end{eqnarray}}

\def\e{\epsilon}

\def\D{\Delta}
\def\del{\delta}
\def\ch{{\mathcal H}}

\def\hcon{\mathrm{H.c.}}

\def\si{\sigma}

\def\o{\omega}
\def\cdag{c^{\dagger}}
\def\da{\downarrow}
\def\ua{\uparrow}
\def\nonum{\nonumber \\}

\def\Om{\Omega}

\def\del{\delta}
\def\mn{\langle n \rangle}

\def\pr{{\sl x. Rev.}\ }
\newcommand{\phrb}[3]{\prb {\bf #1}, #2 (#3)}
\newcommand{\phrl}[3]{\prl {\bf #1}, #2 (#3)}
\begin{document}

\bibliographystyle{prsty}


\title{\Large\bf Pair -- correlations and the survival of superconductivity in and around a
 super-conducting impurity }

\author{Yonatan Dubi}
\affiliation{Physics Department, Ben-Gurion University, Beer Sheva 84105, Israel}
\date{\today}

\begin{abstract} \noindent The problem of the survival of superconductivity in a small super-conducting grain placed 
in a metal substrate is addressed. For this aim the pair correlations and super-conducting gap around and inside
a negative-U impurity in one and two dimensions is calculated, in a discrete
tight-binding model and a continuous model. Using a mean-field decomposition, it is shown
 that finite pairing in the grain develops
 when the system has a degeneracy between successive number of electron pairs, and thus may oscillate as a
 function of the chemical 
potential. For finite pairing in the island, pair correlations in the normal part exhibit a cross-over from 
being long-ranged to exponentially decaying, depending on the strength of interaction in the grain. 
It is shown analytically that there is a minimal island size under-which pairing 
vanishes which is different than that given by Anderson's criterion, and that it scales as a power-law with island size, rather then exponentially
 as in isolated grains. 
\end{abstract}

 \maketitle

\section{Introduction} While superconductivity (SC) on the nano-scale has been a long-standing issue, dating back to the 
seminal work of Anderson \cite{Anderson}, only in the last decade has technological advancement enabled the realization 
of such systems in experiment \cite{RBT}. Since then, manifestations of SC on the nano-meter scale has been observed 
not only in isolated grains \cite{Ralph} or granule on insulating substrate \cite{Bose}, but also 
in inhomogeneous SC thin films, where well-separated SC and normal regions have been observed \cite{STM}. 
In such hybrid systems, the proximity between the SC and normal phases gives rise to novel effects, mainly manifested 
in the local density of states (LDOS), which may be directly measured using scanning tunneling microscopy
 \cite{Kapitulnik}.   

Encouraged by the technological advance, we ask the following question : what would be the properties of 
an ultra-small SC grain placed on a {\sl metallic} (or a doped semi-conducting) substrate ? In such a case,
 would the gap in the grain still 
obey Anderson's criteria \cite{Anderson}, or will the proximity effect yield a new criteria for the destruction 
of SC in such a grain, as seen in, e.g. thin SC layers attached to normal layer \cite{De-Gennes1,MacMillan} ?
 Furthermore, in such systems one expects that the SC properties of the grain would
 be strongly 
affected by the properties of the surrounding metal, and that the proximity to a SC grain would generate 
pair correlations that would impinge 
on the local properties of the metallic area, such as its LDOS \cite{Cooper}. The above question 
is also interesting form a technological point of view, as Josephson arrays fabricated on metallic 
or semi-conducting substrate seem to have large technological potential as nano-electrical devices. 

In order to examine these issues, a minimal model of a single SC grain placed 
in a clean metal matrix is studied, by means of a negative-U Hubbard Hamiltonian in which the
 interactions are confined to a
small region in space (so-called "negative-U impurity"). Applying a Hartree-Fock-Gorkov
 mean-field decomposition \cite{Gorkov} leads to the Bogoliubov-De-Gennes (BdG) Hamiltonian
\cite{De-Gennes}, which serves as a starting point in the calculation. 

The effect of the proximity between the SC grain and normal area is investigated by numerically solving 
a tight-binding BdG Hamiltonian in one and two dimensions. It is found that the pairing in the grain 
is strongly affected by the chemical potential (i.e. density) of the substrate, and that on the normal
area pair-correlations may either be suppressed exponentially away from the impurity or be long-ranged, 
depending on the value of the attractive interaction in the grain. 

The dependence of the gap in the grain on its size is studied using a continuous version of the BdG Hamiltonian. 
Solved analytically, the dependence of the gap on island size is found to diminish as a power-law rather than
 exponential 
(as in an isolated grain \cite{strognin}), and the minimal island size under-which SC vanishes in the grain 
\cite{Spivak} is evaluated,
 and is found to depend on the properties of the substrate. 

\section{The negative-U impurity in the tight-binding model}
Let us start by examining the discrete tight binding model for the
negative-U impurity. The model Hamiltonian is \beq \ch =
-t \sum_{\langle ij \rangle \si} \cdag_{i \si} c_{j \si}-\mu
\sum_{i \si} \cdag_{i \si} c_{i \si} -U \cdag_{0 \uparrow}
\cdag_{0 \downarrow} c_{0 \downarrow} c_{0 \uparrow}~~,
\label{negativeU}\eeq where $t$ is the hopping element, $\mu$ is
the chemical potential and $U>0$ is the attractive interaction,
which is only present in a single site at the origin (the
negative-U impurity). By applying the Hartree-Fock-Gorkov
decomposition \cite{Gorkov} the BdG mean-field
Hamiltonian \cite{De-Gennes} is obtained, \beq \ch =-t \sum_{\langle ij \rangle
\si} \cdag_{i \si} c_{j \si}+\sum_{i \si} (\e_i-\mu)\cdag_{i \si}
c_{i \si}+ \left(\D \cdag_{0 \uparrow} \cdag_{0
\downarrow}+\hcon \right)~~, \label{BdGmeanfield} \eeq where $\D=-U \langle \cdag_{0 \ua} \cdag_{0
\da} \rangle$ is the pairing potential and 
$\e_i
=\del_{i0} \sum_\si |U| \langle \cdag_{0 \si} c_{0 \si} \rangle
/2 = \del_{i0}|U| \langle n_0 \rangle/2 $ is the Hartree shift.  This mean-field
approach is justified by noting that little is known about this system, and
hence a preliminary mean-field treatment is in place.
The importance of maintaining the Hartree shift term, which naturally appears from the derivation 
of the mean-field Hamiltonian, has been discussed and demonstrated in Ref.\onlinecite{ghosal}, where 
a similar decomposition was used to study the properties of a disordered superconducting sample. 

Introducing a Bogoliubov transformation one obtains from the
Hamiltonian of Eq.~\eqref{BdGmeanfield} the BdG equations \cite{De-Gennes} for the
quasi-particle (QP) $u(\mathbf{r}_i)$ and quasi-hole excitations $v(\mathbf{r}_i)$,
\beq \left(
\begin{array}{cc}
 \hat{\xi}+(\e_i-\mu) & \D \del_{i0} \\
 \D^{*} \del_{i0} & -\hat{\xi}- (\e_i-\mu)\\
\end{array}%
\right) \left(
\begin{array}{c}
  u_k(\mathbf{r}_i)  \\
  v_k(\mathbf{r}_i)  \\
\end{array}%
\right)=E_k \left(
\begin{array}{c}
  u_k(\mathbf{r}_i)  \\
  v_k(\mathbf{r}_i)  \\
\end{array}%
\right) ~~. \label{eq:bdg} \eeq In Eq.~(\ref{eq:bdg}) ~ $\hat\xi
u_{k}({\bf r}_i) = -t \sum_{\hat\delta} u_{k}({\bf
r}_i+\hat\delta)$ where $\hat\delta = \pm{\hat{\bf
x}},\pm{\hat{\bf y}}$ and similarly for $v_{k}({\bf r}_i)$, and
the energies are the QP excitation energies $E_k \geq 0$. 
The pairing potential $\D$ and the electron density per site $n_i$ are
to be determined self-consistently in terms of the QP amplitudes
$u(\mathbf{r}_i)$ and $v(\mathbf{r}_i)$ ,
\beq \Delta = |U |\sum_{k}u_{k}(0)v_{k}^{*}(0),
~~~
\langle n_i \rangle = 2 \sum_{k}|v_{k}({\bf r}_{i})|^{2} ~~.
\label {eq:selfc} \eeq

The pairing amplitude $\D$ is finite only on the negative-U
impurity. However, the proximity to the impurity induces
pair-correlations $f_i=\langle \cdag_{i \ua} \cdag_{i \da}
\rangle$ even for $i\neq 0$, that is outside the impurity.  

\section{Results in one and two dimensions}
\begin{figure}[h!] 
\centering
\includegraphics[width=8.5truecm]{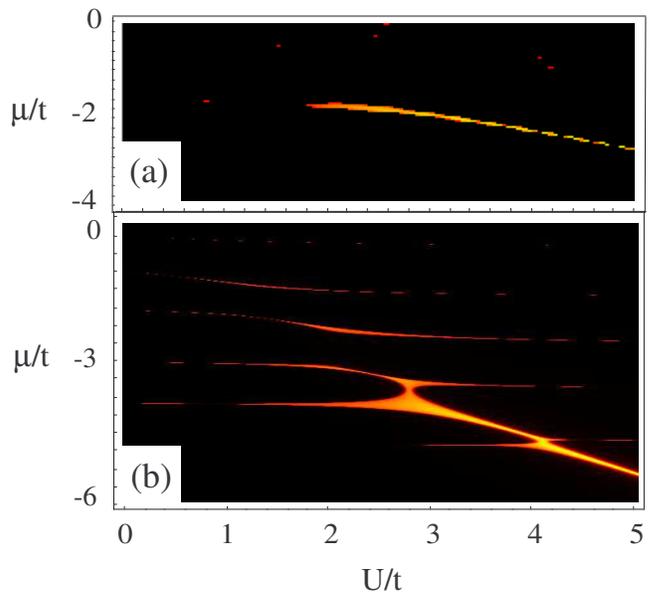} 
\caption{\footnotesize Color-plot of the pairing amplitude in the impurity, $f_0$, as a function 
of chemical potential $\mu$ and interaction strength $U$ for (a) one dimension and (b) two dimensions. Bright area 
indicates large pairing amplitude. }
\label{PhaseDiag1D2D} \end{figure}

The pair-correlations are investigated by solving Eq.~\eqref{eq:bdg}
 numerically and self-consistently for a 1D and 2D metallic substrate, 
from which both the 
densities and pair correlations may be calculated. In Fig.~\ref{PhaseDiag1D2D} the pairing amplitude $f_0$ 
(bright points in Fig.~\ref{PhaseDiag1D2D}) is plotted as a
function of $\mu$ and $U$ for 1D lattice of size $L=49$ and a 2D lattice of size $7 \times 7$ (Fig.~\ref
{PhaseDiag1D2D}(a) and (b), respectively). In the calculation, hard-wall boundary conditions were taken, 
but similar calculation using periodic boundary conditions showed no qualitative change in the results. 
In 1D, finite pairing is only visible 
along the line of constant density $\mn \approx 0$. In 2D stripes of finite pairing appear on 
lines of constant density, which are the boundary lines between sequential even (mean) number of electrons, $N$, in 
the system. It is clear that the phase-space available for finite pairing is much larger in 2D than in 1D. 

\begin{figure}[h!] 
\centering
\includegraphics[width=9.5truecm]{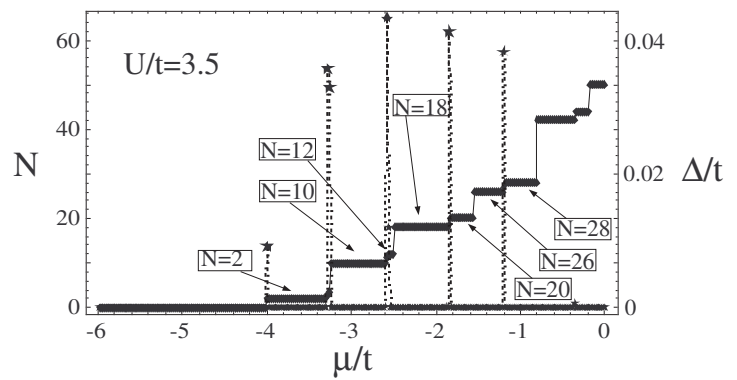} 
\centering
\caption{\footnotesize The pairing potential $\D$ (stars) and mean total electron number $N$ as a function of chemical 
potential for a $13 \times 13$ system with $U/t=3.5$. The pairing potential (stars, right axes) is finite only at 
the transition between plateues of constant even electron number $N$ (diamonds, left axes) }
\label{Nmu2} \end{figure}

In Fig.~\ref{Nmu2} the number of electrons $N$ and the pairing 
potential $\D$ are plotted as a function of 
$\mu$ for a $13 \times 13$ system with $U/t=3.5$. As seen, the pairing potential (stars, right axes)
 is finite only at the transition between plateaus of constant even electron number $N$ (diamonds, left axes).

The reason for this behaviour of the pairing is that the Hamiltonian of Eq.~\eqref{BdGmeanfield} couples between 
states with no electrons and two electrons at the impurity. Thus, the self-consistent pairing is finite only when 
states with $N$ and $N+2$ electrons are degenerate at the Fermi energy. In 1D this only happens 
 when the occupation changes from $N=0$ to $N=2$. In 2D, however, this degeneracy is much more common,
 resulting in many regions of finite pairing in $\mu-U$ phase space, and hence in the oscillatory behaviour 
shown in Fig.~\ref{Nmu2}. We note that while for 3D the computation is numerically demanding and will not be presented here, we expect 
similar behavior, with even more phase space available for SC in the grain. Such a case may be more relevant
 from the experimental side.

\begin{figure}[h!]
\centering
\includegraphics[width=9truecm]{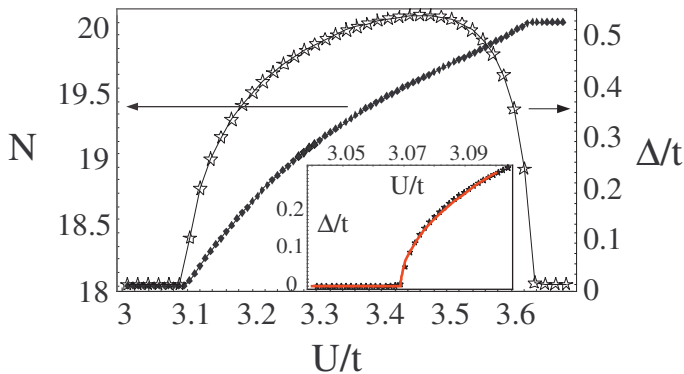}
\centering
\caption{\footnotesize The pairing potential $\D(0)$ and mean total electron number $N$  as a function of interaction
strength $U$ for a $9 \times 9$ lattice with $\mu/t=-2.4$, showing that $\D(U)$ is a non-monotonic function, only finite 
above $U_c$ when $N$ fluctuates between $N=18$ and $N=20$. Inset: the numerical data (stars) in the region $U \simeq 
U_c$ is 
fitted with the function $\D(U) \propto (U/U_c-1)^{x} \Theta(U-U_c)$ (solid line), yielding $x \simeq  1/2$.   }
\label{NDu} \end{figure}

Further insight may be gained by studying the dependence of $\D$ on the interaction strength $U$. Although from Eq.~\eqref
{eq:selfc} it would seem that the two are linearly dependent, this is not the case. Due to the Hartree term, $U$ 
affects both the occupation of the impurity and the energy levels of the system, pushing the system in and out of the 
$N ~-~ N+2$ degeneracy required for finite pairing. This is demonstrated in Fig.~\ref{NDu},
where $\D$ (stars, right axes) and $N$ (diamonds, left axes) are plotted as a function of $U$ for a $9 \times 9$ 
system at $\mu/t=-2.4$. This chemical potential corresponds to a low ($\approx 0.25$) electron filling, which is
 relevant for a semi-conducting substrate.
$\D(U)$ is a non-linear (and non-monotonic) function, only finite above a certain critical interaction $U_c (\mu)$, in 
the transition region of $N$ from $N=18$ to $N=20$. We note that by changing $\mu$ one may find finite pairing at
higher electron densities. However, this is unlikely to occur above half-filling, as the Hartree term will suppress 
the pair function in that case.  

\begin{figure}[h!] 
\centering
\includegraphics[width=8.5truecm]{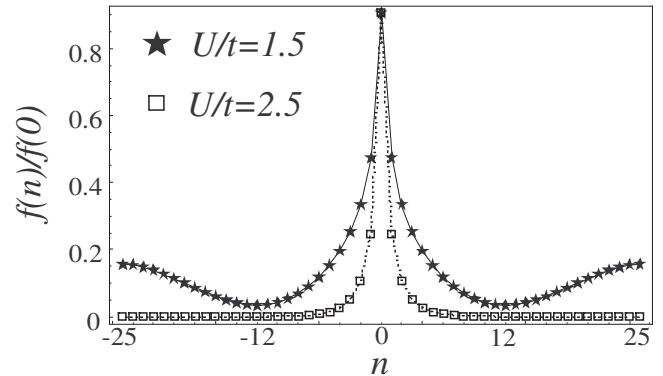} 
\centering
\caption{\footnotesize Spatial structure of the pair correlations $f_i=\langle \cdag_{i \ua} \cdag_{i \da}
\rangle$ in a linear chain of length $L=51$ for two values of interaction, $U/t=1.5$ (stars) and $U/t=2.5$ squares.
 When $U$ lies within the band there are long-range pair-correlations,
 but an exponential decay of the correlations for $U$ outside the energy band.  }
\label{fn} \end{figure}

In the region where the pairing in the SC island is finite, the proximity effect should yield 
pair correlations away from the impurity, $f_i=\langle \cdag_{i \ua} \cdag_{i \da}
\rangle$. In Fig.~\ref{fn} we plot $f_i$ on a chain of length $L=51$ \cite{2D} (normalized to 
unity) for two values of interaction, $U/t=1.5$ (stars), within the energy band, and $U/t=2.5$ (squares), outside the 
band. The chemical potential is adjusted for each value of 
interaction in order to maintain finite pair correlation in the impurity. When $U$ lies outside the band, it is found 
that the hole excitations $v(n)$ become localized, resulting in an 
exponential decay of the pair correlations.
On the other hand,  If $U$ lies within the band the hole excitations are periodic, and generate 
long-range pair correlations. While this effect may be due to the finite size of the normal system, 
these long-ranged correlations may have a crucial effect \cite{us} on the global 
behaviour of a 
system with many negative-U impurities (so-called dilute negative-U model \cite{Spivak,Litak,scalletar}), 
as they determine the effective Josephson coupling between the different impurities.   

\section{Two dimensions - continuous model}The starting point for the following calculation is
 the continuous negative-U Hamiltonian (with $\hbar=1$) \beq \ch
=\sum_{\si} \int d^{2} {\bf r} \Psi^{\dag}_{\si}( {\bf r})
 (-\frac{1}{2m} \nabla^{2}-\mu ) \Psi_{\si}( {\bf r})+\ch_I ~~,
\label{Ham:point}\eeq where $\mu$ is the chemical potential, and
\beq \ch_I= \int d^{2} {\bf r} U({\bf r} )
\Psi^{\dag}_{\uparrow }( {\bf r}) \Psi^{\dag}_{\downarrow }( {\bf
r}) \Psi_{\downarrow }( {\bf r}) \Psi_{\uparrow }( {\bf r}) ~~,
\eeq where $U({\bf r})<0$ is a short-range attractive
electron-electron interaction. The existence of a
negative-U impurity is modelled by limiting the interaction to a finite
island-like region in space \cite{Spivak}, i.e. \beq U({\bf r}) = \left\{
\begin{array}{c}
  -|U|,~~ {\bf r} \in {\bf I } \\
  0,~~ \mathrm{else} \\
\end{array}%
\right. \label{U_r}\eeq where ${\bf I} $ is a disk with radius $a$ around the origin.
\par Using the BdG mean-field decomposition \cite{De-Gennes}, we substitute $\ch_I$ by

\beq \ch_{\D}= \int_{\mathbf{r} \in {I}} d^2 \mathbf{r} ( \Delta (\mathbf{r}) \Psi^{\dag}_{\uparrow}({\bf r})\Psi^{\dag}_{\downarrow}(\mathbf{r}) +\hcon) \quad ~, \eeq where $\D (\mathbf{r}) =|U| \langle
\Psi_{\uparrow}(\mathbf{r})\Psi_{\downarrow}(\mathbf{r})$ is the pairing
potential. The Hamiltonian $\ch_\D$ is now expanded with $a$ being a small parameter
(specifically $k_F a<<1$), \beqa \ch_{\D} &\approx &
\left\{ \D a^2
\Psi^{\dag}_{\uparrow}(0)\Psi^{\dag}_{\downarrow}(0) + \mathrm{
H.c.}+ \right. \\ \nonumber &&  \left.+ \frac{1}{4!}\D^2 a^4
\nabla^2 \left( \Psi^{\dag}_{\uparrow}({\bf
r})\Psi^{\dag}_{\downarrow}({\bf r})+ \hcon \right) \left|_{{\bf
r}=0}  \right. \right\} \quad, \label{point-island} \eeqa where
$\D\equiv\D(0)$. Since $\nabla^2 \psi^{\dag}_{\uparrow}({\bf
r})\psi^{\dag}_{\downarrow}({\bf r}) \left|_{{\bf r}=0} \right.
\sim k_F^2 \Psi^{\dag}_{\uparrow}(0)\Psi^{\dag}_{\downarrow}(0)$,
The second term in Eq.(\ref{point-island}) is smaller than the
first term by a factor of $(k_f a)^2$ and can be neglected, and we
are left with the first term in Eq.(\ref{point-island}) which can
be treated as a local perturbation. Notice that $\D$ is not assumed to be small, as it might not be. 

For typical metals, the condition $k_F a <<1$ yields an island of size of a few nano-meters, 
for which the continuum formulation is inappropriate. For semi-conductors, where the Fermi wave-length may be 
a few orders of magnitude larger than in metals, the continuum limit will still be valid. In semi-conductors
which exhibit true superconductivity, due to the low carrier density the key role is played by inter-valley
processes \cite{Cohen}. In the above model, on the other hand, attractive interactions take place only on the
 impurity and the 
sample as a whole need not become superconducting (on the contrary, it is assumed that it remains normal).
 Thus, local superonducting correlations will probably not be detected by conventional 
transport measurement. However, the manifestation of pairing correlations can still be detected via local measurements of,
 e.g. the LDOS, in which a mini-gap should appear (see Eq.\eqref{DOS1} below).          

\begin{figure}[h!]
\centering
\includegraphics[width=6.5truecm]{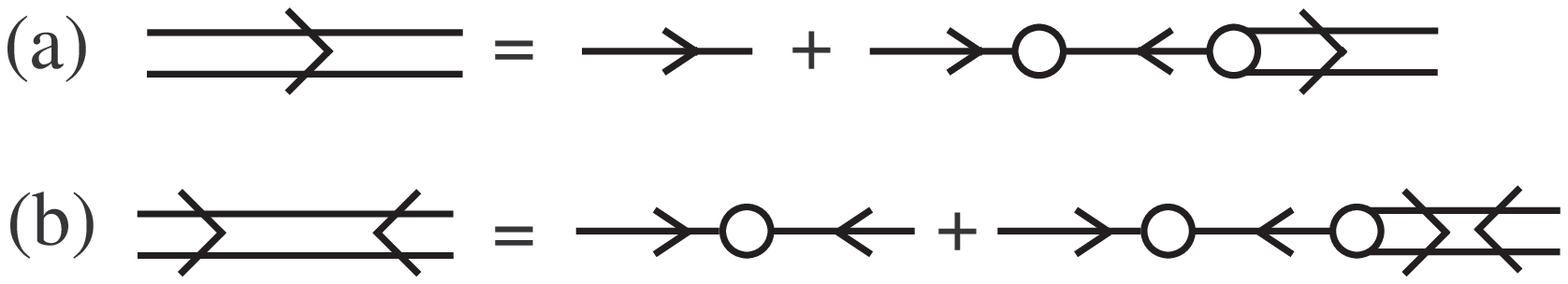}
\centering
\caption{\footnotesize (a) Dyson's equation for the single particle Green's function. (b)
Dyson's equation for the anomalous Green's function.}
\label{Dyson1} \end{figure}

As a first step let us calculate the single particle LDOS in the presence of the impurity, given by
$\rho(\mathbf{r},\o) = -\frac{1}{\pi} \Im G^r (\mathbf{r},\mathbf{r},\o)$
 where $G^r(\mathbf{r},\mathbf{r}',\o)$ is the retarded Green's function, which obeys the Dyson
equation (depicted in Fig.~\ref{Dyson1}) \beqa
G^r(\mathbf{r},\mathbf{r}',\o)&=&g^r(\mathbf{r},\mathbf{r}',\o)+\D^2 a^4 g^r(\mathbf{r},0,\o)\times \nonum
&&~ \times g^r(0,0,-\o) G^r(0,\mathbf{r}',\o)~~, \label{dyson1} \eeqa

Where $g^r(\mathbf{r},\mathbf{r}',\o) =-i \pi \Om J_0 (\sqrt{2m (\mu+\o)}\mid
\mathbf{r}-\mathbf{r}'\mid) $ is the Green's function for a clean 2D system, $J_0$ 
is the zero order Bessel function and $\Om$ is the 2D DOS at the Fermi energy.
 The solution for Eq.~\eqref{dyson1} is \beq
G^r(\mathbf{r},\mathbf{r}',\o)=g^r(\mathbf{r},\mathbf{r}',\o)-\frac{i}{\pi \Om}
\frac{\gamma}{1+\gamma} g^r(\mathbf{r},0,\o)g^r(0,\mathbf{r}',\o)~~,
\label{retardedG} \eeq where $\gamma = \left( \pi \Om \D a^2
\right)^2$. This yields for the DOS
 \beq \rho (\mathbf{r},\o )  \simeq \Om \left( 1-\frac{\gamma}{1+\gamma} J_0(\sqrt{2m (\mu+\o)}\mid \mathbf{r} \mid )^2 \right) ~~,
\label{DOS1} \eeq
 which shows a decrease (so called mini-gap) in the LDOS at the location of the island, and Friedel-like oscillations 
 from the island. The oscillations persist on a length-scale $\sim \sqrt{\Om / \mu}$, independent of 
the pairing potential in the island, which only affects the depth of the minigap. Due to these oscillations 
in the LDOS, one expects that the charge density will re-distribute (and also exhibit oscillations) around 
the SC--semi-conductor interface. One also expects that the density will re-distribute within the
 SC island. However, this effect cannot be probed within the point-island limit (Eq.~\eqref{point-island}), 
and will be discussed in a future study.    

Next we turn to a self consistent calculation of $\D$. Using the Dyson equation
for the anomalous Green's function $F(x,x',t)=\langle {\mathcal T}
\psi^{\dagger}_{\uparrow }(x,0) \psi^{\dagger}_{\downarrow }(x',t)
\rangle$ (Fig.~\ref{Dyson1}(b)) one finds \beq F(x,x',\o)=\frac{\D a^2}{1+\gamma}
g(x,0,\o ) g(0,x',-\o) ~ .
\label{anomalous}\eeq 
The asymptotic properties of Bessel functions
yields $F(x) \sim 1/x$ far from the impurity, independent from 
the value of interaction strength, as expected in a N-SC interface.
 This is in contrast with the exponential decay in the tight-binding model.

The self-consistency equation for the
pairing potential reads (see, e.g., Ref.~\onlinecite{De-Gennes}) \beq
\D(x)=-|U| \int^{\o_D}_{-\o_D}d \o F(x,x,\o)~~,
\label{self-consistency1}\eeq where $\o_D$ is the frequency
cut-off of the interaction. Substituting Eq.~(\ref{anomalous}) yields
\beq \D(x)=-|U| \D a^2 \left( \frac{\gamma}{1+\gamma}
\right) \eta (x)~~, \label{self-consistency2}\eeq 
where $ \eta
(x)= \int^{\o_D}_{-\o_D}d \o g(x,0,\o) g(0,x,-\o)$. At the center
of the impurity $ \eta(0)=-2 (\pi \Om)^2 \o_D$. Inserting this
into Eq.~(\ref{self-consistency2}) results in an algebraic equation
for $\D(0)$, 
\beq
1= 2 (\pi \Om )^2 a^2 |U|\o_D \left( \frac{(\pi \Om a^2 )^2 \D(0)^2}{1+(\pi \Om a^2 )^2 \D (0)^2} \right)~~,
\eeq
which is easily solved to give \beq \D(0)=\frac{1}{\pi
\Om a^2} \left( 2 \pi^2 \Om^2 |U| \o_D a^2-1\right)^{1/2}~~.
\label{Delta}\eeq This self-consistent solution vanishes when
$a=a_I=(2 \pi^2 \Om^2 \o_D |U|)^{-1/2}$, which is the minimal
island area. Let us estimate the minimal island size, $\xi_a \sim a^{1/2}_I$ for a realistic system, 
composed of a Nb island embedded on a 
semi-conducting quantum-well made of Si or GaAs. Taking the effective mass $m^*/m = 0.98$ and
 $m^*/m = 0.063$ for Si and GaAs respectively,\cite{handbook} one can estimate the 2D-DOS in the quantum well.
 Taking for Nb
$T_c=9.26K$ and $ \theta_D=275K$,\cite{AM} one finds that for the Nb/Si hybrid the minimal
 island radius is $\xi_I \sim 80$nm, 
and for the Nb/GaAs system $\xi_I \sim 20$nm. Both these lengths are still in the point-island regime, since the 
Fermi wave-length may be an order of magnitude larger for such quantum wells.

Eq.~(\ref{Delta}) also supplies us with a
dependence of $\D$ on the island size and interaction strength. In the inset 
of Fig.~\ref{NDu} we plot a fit of the numerical data in the region $U\simeq U_c$ (stars) 
to a function of the form $\D(U) \propto (U/U_c-1)^{x} \Theta(U-U_c)$ (solid line), as in Eq.~\eqref{Delta}. 
The fit yields the exponent $x=0.5002$, in good agreement with the continuous model. 

In the definition of the model (Eq.\eqref{Ham:point}-\eqref{U_r}) we have neglected the boundary conditions 
on the normal-SC interface. Omitting the boundary effect is hard to justify a-priory, especially when 
the size of the SC grain is smaller than the SC coherence length. Accounting for the boundary should be accounted 
for by solving Eqs.\eqref{anomalous}-\eqref{self-consistency2} with an additional constraint $\D(a)=0$. However, 
the comparison between the numerical calculation (in which the boundary condition 
are inherently implemented) and the analytical result (inset of Fig.~\ref{NDu}) shows a striking equivalence between
 them. This indicates that neglecting the boundary effect merely results in a quantitative modification.
It does not change the qualitative behavior, which is mainly characterized by the power-law dependence
specified in Eq.\eqref{Delta}.

\section{Discussion}
One main feature of the result shown in Eq.~\eqref{Delta} is that, in contrast to previous works on the proximity
effect \cite{De-Gennes1,MacMillan,Spivak}, the length-scale is not the usual superconducting coherence length $\xi$, 
but rather a new length scale $a_I= (2 \pi^2 \Om^2 \o_D |U|)^{-1/2}$. In order to understand the origin of this new
length scale we cast the criterion for the vanishing of SC correlations in the island given by Eq.~\eqref{Delta} to the 
form 
\beq
\frac{1}{\sqrt{2}}\left( \frac{|U|}{\o_D}\right)^{1/2} 2 \pi a^2 \Om \o_D =1~~.\label{condition}
\eeq
The factor $2 \pi a^2 \Om \o_D$ is nothing but the number of electrons with energy in the range $\o_D$ 
within the island, and thus the condition turns out to be a continuous version of the Anderson criteria, 
which states that SC vanishes in a grain once the number of pairs, roughly given by $\D/\del$ where $\del$ is the 
level spacing, becomes less then unity. However, in the above model there is no discreteness of energy 
levels. Rather, the number of pairs in restricted due to the finite region to which 
the interaction is limited to. It is also clear from this argument why $\xi$ does
not play a role in this system, as $\xi$ indicates the existence of a region where superconductivity is developed 
to its bulk value, which is not the case here. 

Yet another way of understanding the result of Eq.~\eqref{Delta} is to note that in a SC--semi-conductor junction, 
one expects that due to the low density on the normal side, the suppression of pair correlations in the SC
 (due to the proximity effect) 
will no longer be on a length scale $\xi$ as in a SC-metal junction, but rather a new length scale, which in the
 point-island approximation corresponds to $\xi_I \approx a^{1/2}_I$. Thus, if the SC island is smaller than 
the length-scale on which pair correlations are suppressed, SC in the island will vanish. One also expects that
upon increasing the island size (beyond the point-island approximation) this new length-scale will change. This 
problem is beyond the scope of the present work and will be addressed in a future study.

In Eq.~\ref{Delta}, $\D$ has a power-law dependence on the grain size. This 
is in contrast to the case of an isolated grain, where an exponential dependence on size is predicted \cite{strognin}.
Furthermore, due to the re-normalization of electron number and the dependence of critical island size on the
 DOS, the critical size may be either larger or smaller 
than given by Anderson's criteria. This may affect the possibility of fabricating devices made from SC grains 
embedded on a metallic matrix. This effect may be tested experimentally by, e.g., varying the DOS of the metallic 
substrate by changing its carrier density (by gating the sample, for instance).

More intuition on the existence of a critical island size or interaction strength may be obtained
by noting the similarity between Eq.~\ref{Delta} and Anderson's criteria for the existence of a magnetic impurity 
in a metal \cite{Anderson2}. This similarity, along with the Friedel-like oscillations of Eq.~\ref{DOS1},
 implies that 
the Negative-U impurity is screened by the free electron gas, in an analogous way to the screening of a magnetic 
impurity. It would thus be intriguing to investigate the possibility of the formation of an effect equivalent
 to the charge Kondo effect \cite{chargeKondo}, resulting from the presence of embedded SC grains.  

We conclude by noting that these results may be tested experimentally, by planting superconducting impurities
on a metallic substrate and measuring the local gap, in a similar way to that of Ref.~\onlinecite{Bose}.
 Another system in which our results may be valid is a SC grain 
strongly coupled to matching leads (i.e. SC Aluminium grain and normal Aluminium leads), where the
 existence of SC may
be verified as a function of grain size. While this may be experimentally challenging,
 it may be achieved by, for instance, planting magnetic impurities in the leads. 
The dependence of the SC gap on the properties of the substrate,as seen in
Fig.~\ref{Nmu2} for instance, may be tested by changing the density on the metallic substrate.   

The author acknowledges fruitful discussions with Y. Meir and Y. Avishai, and is grateful to T. Aono 
for helpful discussions and for carefully reading the manuscript. 
This research has been funded by the ISF.


\begin{references}
\bibitem{2D}
A similar calculation was performed for the 2D lattice, yielding a similar result.
\end{references}
\end{document}